\documentclass[12pt]{article}
\usepackage{bbm}
\newcommand{\be}{\begin{equation}}
\newcommand{\ee}{\end{equation}}
\newcommand{\ba}{\begin{eqnarray}}
\newcommand{\ea}{\end{eqnarray}}
\newcommand{\no}{\nonumber\\}
\begin{document}
\title{\normalsize \hfill UWThPh-2003-23 \\[1cm] \LARGE
Maximal atmospheric neutrino mixing \\
and the small ratio of muon to tau mass}
\author{Walter Grimus\thanks{E-mail: walter.grimus@univie.ac.at} \\
\setcounter{footnote}{3}
\small Institut f\"ur Theoretische Physik, Universit\"at Wien \\
\small Boltzmanngasse 5, A--1090 Wien, Austria \\*[3.6mm]
Lu\'{\i}s Lavoura\thanks{E-mail: balio@cfif.ist.utl.pt} \\
\small Universidade T\'ecnica de Lisboa \\
\small Centro de F\'\i sica das Interac\c c\~oes Fundamentais \\
\small Instituto Superior T\'ecnico, P--1049-001 Lisboa, Portugal \\*[4.6mm] }

\date{27 October 2003}

\maketitle

\begin{abstract}
We discuss the problem of
the small ratio of muon mass to tau mass
in a class of seesaw models where maximal atmospheric neutrino mixing
is enforced through a $\mu$--$\tau$ interchange symmetry.
We introduce into those models
an additional symmetry
$K$ such that $m_\mu = 0$ in the case of exact
$K$ invariance.
The symmetry
$K$ may be softly broken in the Higgs potential,
and one thus achieves $m_\mu \ll m_\tau$ in a technically natural way.
We speculate on a wider applicability of this mechanism.
\end{abstract}

\newpage

\section{Introduction} \label{Introduction}

Lepton mixing seems to be an established fact now---for reviews see,
for instance,
Ref.~\cite{reviews}.
The solar neutrino mixing has turned out to be large but non-maximal,
whereas it is likely that the atmospheric mixing angle is maximal,
i.e.\ $\pi / 4$ or close to that value.

On the theoretical side,
a popular way of generating small neutrino masses
is the seesaw mechanism \cite{seesaw}.
It has three sources of lepton mixing:
the charged-lepton mass matrix $M_\ell$,
the Dirac mass matrix $M_D$
linking the left-handed neutrinos $\nu_L$
to the right-handed neutrinos $\nu_R$,
and the Majorana mass matrix $M_R$ of the right-handed neutrinos;
the mass Lagrangian is
\be
\mathcal{L}_\mathrm{mass} =
- \bar \ell_R M_\ell \ell_L
- \bar \nu_R M_D \nu_L
- \frac{1}{2}\, \bar \nu_R M_R C \bar \nu_R^T +
\mbox{H.c.}
\label{MR}
\ee
The mass matrix of the light neutrinos is then given by
\be
\mathcal{M}_\nu = - M_D^T M_R^{-1} M_D\, .
\ee 

Some models have been proposed in the literature \cite{ma,GL01,GL03,GLcp}
which account for maximal atmospheric neutrino mixing
by invoking non-abelian symmetries such that $M_\ell$ and $M_D$
are simultaneously diagonal.
In this case the mass matrix $M_R$
is the sole source of neutrino mixing.
An interesting common feature of the models of
Refs.~\cite{GL01,GL03,GLcp} 
is that they need a minimum of three Higgs doublets
in order to have enough freedom to accommodate all lepton masses.

In this paper we investigate the ``$\mathbbm{Z}_2$ model'' of Ref.~\cite{GL01},
the ``$D_4$ model'' of Ref.~\cite{GL03},
and the ``$CP$ model'' of Ref.~\cite{GLcp}.
The models of Refs.~\cite{GL01,GL03}
have non-abelian horizontal-symmetry groups.
The model of Ref.~\cite{GLcp}
has a non-standard $CP$ symmetry \cite{scott}---for the general idea
of non-standard $CP$ transformations see Ref.~\cite{ecker}---instead of
a certain $\mathbbm{Z}_2$ contained in the symmetry groups
of Refs.~\cite{GL01,GL03}.
The symmetries defining the models of Refs.~\cite{GL01,GL03,GLcp}
will be described in the respective sections of this paper;
there,
their scalar sectors are also discussed. 

In all those models the masses of the charged leptons
are free and must be adjusted by finetuning. 
The finetuning problem of adjusting the small ratio of muon mass ($m_\mu$)
to tau mass ($m_\tau$) is more complex than in the Standard Model,
as we shall see below.
In Ref.~\cite{GL03} we have proposed a solution to this problem
by introducing an additional symmetry operation
$K$---but no additional
fields---which restricts the coupling constants of the model
in such a way that $m_\mu = 0$ with exact
$K$ invariance.\footnote{In the paper of Ref.~\cite{GL03} we have used
$T$ instead of $K$ for denoting the extra symmetry which leads to $m_\mu = 0$.
Since $T$ is commonly used in physics for the time-reversal symmetry,
in this paper and in Ref.~\cite{GLcp} we have switched the notation to $K$.}
Through soft
$K$ breaking a non-zero ratio $m_\mu/m_\tau$ is then achieved,
and this solves the finetuning problem in a technically natural way. 
The problem of reconciling maximal atmospheric neutrino mixing with a
small ratio $m_\mu/m_\tau$ was also addressed in Ref.~\cite{seidl} in
a class of models different from the ones discussed here.

The purpose of this
paper is twofold:
firstly,
we present the detailed proof of the existence of this mechanism
and work out the principles on which it is based;
secondly,
we show that it may operate not only in the model
of Ref.~\cite{GL03} but also in those of Refs.~\cite{GL01,GLcp}.
It will become evident to which class of models the symmetry
$K$ can be applied.

In Section~\ref{Yukawa} we describe the multiplets and Yukawa
Lagrangians of the models of Refs.~\cite{GL01,GL03,GLcp},
we point out the finetuning problem for $m_\mu \ll m_\tau$,
and we introduce the symmetry
$K$,
which reduces the finetuning to a problem of
achieving nearly equal vacuum expectation values (VEVs)
for two Higgs doublets. 
The latter problem is addressed by investigating the effect of
$K$ on the Higgs potentials of the $\mathbbm{Z}_2$,
$D_4$,
and $CP$ models in Sections~\ref{Z2model},
\ref{D4model},
and~\ref{CPmodel},
respectively.
Section~\ref{summary} contains a summary.

\section{The Yukawa Lagrangians and $K$} \label{Yukawa}

In Refs.~\cite{GL01,GL03}
there are the following Yukawa couplings in the lepton sector:
\ba
\mathcal{L}_\mathrm{Y} & = & 
- y_1 \bar D_e \nu_{eR} \tilde\phi_1  
- y_2 \left( \bar D_\mu \nu_{\mu R} + \bar D_\tau \nu_{\tau R} \right)
\tilde\phi_1 
\nonumber \\ && \label{L}
- y_3 \bar D_e e_R \phi_1
- y_4 \left( \bar D_\mu \mu_R + \bar D_\tau \tau_R \right) \phi_2
- y_5 \left( \bar D_\mu \mu_R - \bar D_\tau \tau_R \right) \phi_3 
+ \mbox{H.c.} 
\ea
The constants $y_i$ ($i=1,\ldots,5$) are in general complex.
Denoting the VEV of $\phi_j^0$ ($j=1,2,3$) by $v_j \left/ \sqrt{2} \right.$, 
the muon and tau masses are given by
\be
m_\mu  = \frac{1}{\sqrt{2}} \left| y_4 v_2 + y_5 v_3 \right|,
\quad
m_\tau = \frac{1}{\sqrt{2}} \left| y_4 v_2 - y_5 v_3 \right|,
\ee
respectively.

A variant of the Lagrangian of Eq.~(\ref{L}),
namely
\ba
\mathcal{L}'_\mathrm{Y} & = & 
- y_1 \bar D_e \nu_{eR} \tilde\phi_1  
- \left( y_2 \bar D_\mu \nu_{\mu R} + y_2^* \bar D_\tau \nu_{\tau R} \right)
\tilde\phi_1 
\nonumber \\ && \label{L'}
- y_3 \bar D_e e_R \phi_1
- \left( y_4 \bar D_\mu \mu_R +  y_4^* \bar D_\tau \tau_R \right) \phi_2
- \left( y_5 \bar D_\mu \mu_R -  y_5^* \bar D_\tau \tau_R \right) \phi_3
+ \mbox{H.c.}\, ,
\ea
is found in the model of Ref.~\cite{GLcp}.
Here,
$y_1$ and $y_3$ are real,
but the remaining $y_i$ are in general complex; 
the muon and tau masses are given by
\be
m_\mu  = \frac{1}{\sqrt{2}} \left| y_4 v_2 + y_5 v_3 \right|,
\quad
m_\tau = \frac{1}{\sqrt{2}} \left| y_4^* v_2 - y_5^* v_3 \right|,
\ee
respectively.

In all three models under discussion there is an ``auxiliary''
symmetry of the $\mathbbm{Z}_2$ type:
\be
\mathbbm{Z}_2^{(\mathrm{aux})}: \;
\nu_{eR},\: \nu_{\mu R},\: 
\nu_{\tau R},\: \phi_1,\:
e_R\, \: \mbox{change sign.}
\label{Z2aux}
\ee
It is spontaneously broken by $v_1$
and it restricts the couplings of the Higgs doublets $\phi_j$
as seen in the Yukawa Lagrangians of Eqs.~(\ref{L}) and~(\ref{L'}).

One must use finetuning for obtaining $m_\mu \ll m_\tau$.
In the case of the Lagrangians of Eqs.~(\ref{L}) and~(\ref{L'})
one does not simply have to choose one Yukawa coupling to be small,
like in the Standard Model,
rather one has to choose two products of unrelated quantities---one
Yukawa coupling and one VEV---such that in $m_\mu$
those two products nearly cancel.
In order to soften the amount of finetuning
we have proposed in Ref.~\cite{GL03}
to add to the model of that paper the symmetry 
\be
K: \quad \mu_R \to -\mu_R\, , \quad \phi_2 \leftrightarrow  \phi_3\, .
\label{T}
\ee
That symmetry leads to
\be
y_4 = - y_5
\label{yy}
\ee
in both the Lagrangians of Eqs.~(\ref{L}) and (\ref{L'}). 
In this way the finetuning is confined to the VEVs:
\be
\frac{m_\mu}{m_\tau} = \left| \frac{v_2 - v_3}{v_2 + v_3} \right|.
\label{ratio}
\ee
The symmetry
$K$ also has effects upon the scalar potential $V$,
which we now turn to investigate.
In this task,
there are some differences among the scalar sectors
and among the symmetries of the models of Refs.~\cite{GL01,GL03,GLcp}
which must be taken into consideration.

\section{The $\mathbbm{Z}_2$ model} \label{Z2model}

The scalar sector of the model of Ref.~\cite{GL01}
consists solely of the three Higgs doublets $\phi_j$.
Its symmetries are the following:
\begin{itemize}
\item the three groups $U(1)_{L_\alpha}$ ($\alpha = e,\mu,\tau$)
associated with the lepton numbers $L_\alpha$,
which are softly broken by the Majorana mass terms
of the right-handed neutrino singlets,
i.e.\ by the matrix $M_R$;
\item a $\mathbbm{Z}_2$-type symmetry given by 
\be
\mathbbm{Z}_2^{(\mathrm{tr})}: \;
D_\mu \leftrightarrow D_\tau\, , \: \mu_R \leftrightarrow \tau_R\, , \:
\nu_{\mu R} \leftrightarrow \nu_{\tau R}\, , \: \phi_3 \to - \phi_3\, ,
\label{Z2tr}
\ee
which transposes the second and third families
and is spontaneously broken by $v_3$;
\item the symmetry $\mathbbm{Z}_2^{(\mathrm{aux})}$ of Eq.~(\ref{Z2aux}).
\end{itemize}
Because of the two $\mathbbm{Z}_2$-type symmetries
the Higgs potential $V$ of this model is 
invariant under the three independent sign changes 
\be
\phi_j \to -\phi_j \,, \quad \phi_{j'} \to \phi_{j'} \quad (j' \neq j) \,,
\label{signs}
\ee
where $j, j' = 1,2,3$. 
Therefore,
in every term of $V$ each Higgs doublet
can occur only either two or four times. 

We define the Higgs potential $V_\phi$ as the polynomial of order 4 
in the $\phi_j$ obeying the symmetries of Eqs.~(\ref{T}) and (\ref{signs}).
\emph{Soft breaking} of the symmetry
$K$ of Eq.~(\ref{T}) 
is achieved by adding to $V_\phi$ the unique term
\be
V_\mathrm{soft} = \mu_\mathrm{soft}
\left( \phi_2^\dagger \phi_2 - \phi_3^\dagger \phi_3 \right).
\label{soft}
\ee
We want to show the following \cite{GL03}:
\begin{enumerate}
\renewcommand{\labelenumi}{(\alph{enumi})}
\item Under certain conditions on the coupling constants,
the minimum of $V_\phi$ fulfills 
\be
v_2 = v_3 \,.
\label{=}
\ee
Equation~(\ref{ratio}) then gives $m_\mu = 0$.
\item The full potential $V = V_\phi + V_\mathrm{soft}$  
leads to a non-vanishing $m_\mu$.
For small $\mu_\mathrm{soft}$ the ratio $m_\mu/m_\tau$ is small
in a technically natural way.
\end{enumerate}
\paragraph{Point (a):} $V_\phi$ is given by
\ba
V_\phi
&=& -\mu_1 \phi_1^\dagger \phi_1
- \mu_2 \left( \phi_2^\dagger \phi_2 + \phi_3^\dagger \phi_3 \right)
\no & &
+ \lambda_1 \left( \phi_1^\dagger \phi_1 \right)^2
+ \lambda_2 \left[ \left( \phi_2^\dagger \phi_2 \right)^2
+ \left( \phi_3^\dagger \phi_3 \right)^2 \right]
\no & &
+ \lambda_3 \left( \phi_1^\dagger \phi_1 \right) 
\left( \phi_2^\dagger \phi_2 + \phi_3^\dagger \phi_3 \right)
+ \lambda_4 \left( \phi_2^\dagger \phi_2 \right)
\left( \phi_3^\dagger \phi_3 \right)
\no & &
+ \lambda_5 \left[
\left( \phi_1^\dagger \phi_2 \right) \left( \phi_2^\dagger \phi_1 \right)
+ \left( \phi_1^\dagger \phi_3 \right) \left( \phi_3^\dagger \phi_1 \right)
\right]
+ \lambda_6
\left( \phi_2^\dagger \phi_3 \right) \left( \phi_3^\dagger \phi_2 \right)
\no & &
+ \lambda_7 \left[ \left( \phi_2^\dagger \phi_3 \right)^2 + 
\left( \phi_3^\dagger \phi_2 \right)^2 \right]
+ \lambda_8 \left[ \left( \phi_1^\dagger \phi_2 \right)^2
+ \left( \phi_1^\dagger \phi_3 \right)^2 \right]
+ \lambda_8^\ast \left[ \left( \phi_2^\dagger \phi_1 \right)^2
+ \left( \phi_3^\dagger \phi_1 \right)^2 \right].
\no & &
\label{Vphi}
\ea
All the coupling constants in $V_\phi$,
except for $\lambda_8$,
are real.
We replace the Higgs doublets in $V_\phi$ by their VEVs,
parameterized as
\be
\frac{v_1}{\sqrt{2}} = u_1\, ,
\quad
\frac{v_2}{\sqrt{2}} = u e^ {i\alpha} \cos{\sigma}\, ,
\quad
\frac{v_3}{\sqrt{2}} = u e^ {i\beta} \sin{\sigma}\, ,
\label{VEV1}
\ee
where,
without loss of generality,
$u_1$ and $u$ are positive and $\sigma$ belongs to the first quadrant.
Note that $\sqrt{2 \left( u_1^2 + u^2 \right)} \simeq 246\, \mathrm{GeV}$
represents the electroweak scale. 
In this way we obtain the function 
\ba
F_\phi \equiv 
\left\langle 0 \left| V_\phi \right| 0 \right\rangle &=&
- \mu_1 u_1^2 - \mu_2 u^2
+ \lambda_1 u_1^4 + \lambda_2 u^4
+ \left( \lambda_3 + \lambda_5 \right) u_1^2 u^2
\nonumber \\ & &
+ \left[ \tilde \lambda
- 4 \lambda_7 \sin^2{\left( \alpha - \beta \right)} \right] 
u^4 \cos^2{\sigma} \sin^2{\sigma}
\no & &
+ 2 \left|\lambda_8 \right| u_1^2 u^2 
\left[ \cos^2 \sigma\, \cos \left( \epsilon + 2\alpha \right)
+ \sin^2 \sigma\, \cos \left( \epsilon + 2 \beta \right) \right],
\label{Fphi} 
\ea
where $\tilde \lambda \equiv - 2 \lambda_2 + \lambda_4 + \lambda_6
+ 2 \lambda_7$ and $\epsilon \equiv \arg{\lambda_8}$.
In searching for the minimum of $F_\phi$ in terms of $\sigma$,
$\alpha$,
and $\beta$ we make the following simple observation:
if
\be
\tilde \lambda < 0 \quad \mbox{and} \quad \lambda_7 < 0 \,,
\label{sufficient}
\ee
then the minimum of $F_\phi$ has
\be
\sigma = \frac{\pi}{4}\, ,
\quad
\alpha = \beta = \frac{\pi - \epsilon}{2}\, ,
\label{min}
\ee
i.e.
\be
\label{v2v3}
v_2 = v_3 = u e^{i \left. \left( \pi - \epsilon \right) \right/ 2}\, .
\ee
With Eq.~(\ref{v2v3}) one obtains $m_\mu = 0$,
which is close to the real situation
if one takes into account that $m_\mu \ll m_\tau$.
Equation~(\ref{sufficient}) formulates sufficient conditions
for the minimum of $V_\phi$ to obey the relation (\ref{v2v3}).
\paragraph{Point (b):} Now we take into account $V_\mathrm{soft}$,
which gives
\be
\langle 0 | V_\mathrm{soft} | 0 \rangle = 
\mu_\mathrm{soft} u^2 \cos{2 \sigma} \,.
\label{softVEV}
\ee
We must minimize 
$\left\langle 0 \left| V_\phi + V_\mathrm{soft} \right| 0 \right\rangle$.
Since the new term in Eq.~(\ref{softVEV})
does not contain the phases $\alpha$ and $\beta$,
the minimum is again at $\alpha = \beta = \left( \pi - \epsilon \right) / 2$.
Then,
the second line of Eq.~(\ref{Fphi})
together with the term stemming from $V_\mathrm{soft}$ lead to
\be
\cos 2\sigma = \frac{2 \mu_\mathrm{soft}}{\tilde \lambda u^2}\, .
\ee
It is obvious that,
if $\left| \mu_\mathrm{soft} \right|$ is sufficiently smaller than
$u^2$,
the angle $\sigma$ will still be close to $\pi / 4$. 
Using Eqs.~(\ref{ratio}) and (\ref{VEV1}),
we obtain 
\be
\frac{m_\mu}{m_\tau} = 
\frac{| \cos 2\sigma |}{1 + \sqrt{1 - \cos^2 2\sigma}} \simeq 
\left| \frac{\mu_\mathrm{soft}}{\tilde \lambda u^2} \right|,
\label{keyrq}
\ee
showing explicitly that a small $\mu_\mathrm{soft}$
leads to a small $m_\mu/m_\tau$.

Let us elaborate on the magnitude of $\mu_\mathrm{soft}$. 
The Higgs-potential coupling $\tilde \lambda$ could be of order $0.1$.
Furthermore, let us choose, for instance, $u = 172$ GeV. 
Then it turns out that $u_1 \simeq 26$ GeV, which is much smaller than
$u$; this is reasonable since $u_1$ 
is responsible both for the electron mass and for the neutrino
masses.
From Eq.~(\ref{keyrq}) one then obtains
$\left| \mu_\mathrm{soft} \right| \sim 176$ GeV$^2$. 
Thus, $\left| \mu_\mathrm{soft} \right|$ is much smaller than $u^2$,
yet it
may well be of the order of magnitude or even larger than $u_1^2$.
On the other hand,
if $u_1$ is comparable to $u$, then we shall have 
$\left| \mu_\mathrm{soft} \right|$ much smaller than both $u^2$ and $u_1^2$,
but its order of magnitude will not change with respect
to the numerical example above.

\section{The $D_4$ model} \label{D4model}

The model of Ref.~\cite{GL03}
is based on the horizontal-symmetry group $D_4$.
It has the same gauge multiplets of the model of the previous section,
plus two real scalar gauge singlets $\chi_k$ ($k=1,2$).
The symmetries are the following:
the horizontal group $D_4$ is generated by the
$\mathbbm{Z}_2^{(\mathrm{tr})}$ of Eq.~(\ref{Z2tr}),
supplemented by $\chi_1 \leftrightarrow \chi_2$,
and by an additional $\mathbbm{Z}_2$-type symmetry 
\be
\mathbbm{Z}_2^{(\tau)}: \;
D_\tau,\ \tau_R,\ \nu_{\tau R},\
\chi_2\ \mbox{change sign.}
\label{Z2tau}
\ee
Besides $D_4$,
there is also the symmetry $\mathbbm{Z}_2^{(\mathrm{aux})}$
of Eq.~(\ref{Z2aux}). 
Both $D_4$ and $\mathbbm{Z}_2^{(\mathrm{aux})}$ are spontaneously broken.
We furthermore impose the symmetry
$K$ of Eq.~(\ref{T}).

The full potential has the structure
\be
V = V_\chi + V_{\chi\phi} + V_\phi + V_\mathrm{soft}\, ,
\label{V}
\ee
where $V_\chi$ contains only the fields $\chi_k$
whereas the terms in $V_{\chi\phi}$
contain both the $\chi_k$ and the $\phi_j$.
Because $\mathbbm{Z}_2^{(\mathrm{aux})}$ and $\mathbbm{Z}_2^{(\mathrm{tr})}$
hold in this case as well,
$V_\phi$ and $V_\mathrm{soft}$ are the same as in the previous section.  
We will show that---like for $V_\phi$ under point (a) in the previous
section---there is a minimum of  
$V_\chi + V_{\chi\phi} + V_\phi$ for which Eq.~(\ref{=}) holds.
Then the procedure of point (b) works in the present model as well. 

The potentials $V_\chi$ and $V_{\chi \phi}$ are given by 
\ba
V_\chi &=&  - \mu \left( \chi_1^2 + \chi_2^2 \right) 
+ \lambda \left( \chi_1^2 + \chi_2^2 \right)^2
+ \lambda^\prime \left( \chi_1^2 - \chi_2^2 \right)^2,
\label{pot1} \\
V_{\chi\phi} &=& 
\left( \chi_1^2 + \chi_2^2 \right)
\left[ \rho_1 \phi_1^\dagger \phi_1
+ \rho_2 \left( \phi_2^\dagger \phi_2 + \phi_3^\dagger \phi_3 \right) \right]
+ \eta \left( \chi_1^2- \chi_2^2 \right)
\left( \phi_2^\dagger \phi_3 + \phi_3^\dagger \phi_2 \right).
\label{pot12}
\ea
In Eqs.~(\ref{pot1}) and (\ref{pot12}) all the coupling constants are real. 
We parameterize the VEVs of the $\chi_k$ in the following way \cite{GL03}:
\be
\left\langle 0 \left| \chi_1 \right| 0 \right\rangle = W \cos{\gamma}\,,
\quad
\left\langle 0 \left| \chi_2 \right| 0 \right\rangle = W \sin{\gamma}\,.
\ee
We then obtain 
\ba
\left\langle 0 \left| V_\chi + V_{\chi \phi} \right| 0 \right\rangle
&=& - \mu W^2 + \lambda W^4 + W^2 \left( \rho_1 u_1^2 + \rho_2 u^2 \right)
\no & & 
+ 2 \eta W^2 u^2 \cos{2 \gamma} \cos{\sigma} \sin{\sigma}
\cos{\left( \alpha - \beta \right)}
+ \lambda^\prime W^4 \cos^2{2 \gamma}\, .
\label{chiphi}
\ea
The minimum of $\gamma$ is found at 
\be
\label{gamma}
\cos{2 \gamma} = - \frac{\eta u^2 \cos{\sigma} \sin{\sigma}
\cos{\left( \alpha - \beta \right)}}{\lambda^\prime W^2}\, ,
\ee
provided $\lambda^\prime > 0$
and the quantity in the right-hand side of Eq.~(\ref{gamma})
is smaller than unity in modulus.\footnote{In the model
of Ref.~\cite{GL03} we need $W \gg u$ anyway.}
Inserting the result of Eq.~(\ref{gamma}) into Eq.~(\ref{chiphi})
we arrive at 
\ba
\left\langle 0 \left| V_\chi + V_{\chi \phi} \right| 0 \right\rangle
&=& - \mu W^2 + \lambda W^4
+ W^2 \left( \rho_1 u_1^2 + \rho_2 u^2 \right)
\no & &
- \frac{\eta^2}{\lambda^\prime}\, u^4 \cos^2{\sigma} \sin^2{\sigma}
\cos^2{\left( \alpha - \beta \right)}\, .
\ea
Adding this expression to $F_\phi$ in Eq.~(\ref{Fphi}), we see that
this amounts to the replacements 
$\tilde \lambda \to \tilde \lambda - \eta^2/\lambda'$ and
$4\lambda_7 \to 4\lambda_7 - \eta^2/\lambda'$ in Eq.~(\ref{Fphi}). 
Then the minimum of $V_\chi + V_{\chi\phi} + V_\phi$ with respect to $\sigma$,
$\alpha$,
and $\beta$ is not changed compared to Eq.~(\ref{min});
we have again $v_2 = v_3$ and,
therefore,
$m_\mu = 0$.
As before,
breaking the symmetry
$K$ softly through $V_\mathrm{soft}$
one achieves a non-zero muon mass.
The muon mass is small in a technically natural way 
by requiring only one parameter,
$\mu_{\rm soft}$,
to be small.

\section{The $CP$ model} \label{CPmodel}

The model of Ref.~\cite{GLcp} has the same scalar multiplets
and nearly the same symmetries
as the $\mathbbm{Z}_2$ model of Section~\ref{Z2model};
the only difference is
that the $\mathbbm{Z}_2^{(\mathrm{tr})}$ of Eq.~(\ref{Z2tr})
is replaced by the non-standard $CP$ symmetry 
\be
\renewcommand{\arraystretch}{1.2}
\begin{array}{rcl}
D_\alpha &\to& i S_{\alpha \beta} \gamma^0 C \bar D_\beta^T\, , \\
\alpha_R &\to& i S_{\alpha \beta} \gamma^0 C \bar \beta_R^T\, , \\
\nu_{\alpha R} &\to& i S_{\alpha \beta} \gamma^0 C \bar \nu_{\beta R}^T\, , \\
\phi_{1,2} &\to& \phi_{1,2}^\ast\, , \\
\phi_3 &\to& - \phi_3^\ast\, , 
\end{array}
\quad \mbox{with} \quad
\renewcommand{\arraystretch}{1}
S = \left( \begin{array}{ccc} 
1 & 0 & 0 \\ 0 & 0 & 1 \\ 0 & 1 & 0 \end{array} \right).
\label{cp}
\ee
Here,
$\alpha, \beta = e, \mu, \tau$.

This replacement not only leads to the Yukawa Lagrangian
of Eq.~(\ref{L'}),
slightly different from the one of Eq.~(\ref{L}),
but also allows for a more general Higgs potential,
due to the complex conjugation in the transformation of the Higgs doublets
in Eq.~(\ref{cp}). 
One can show that invariance under the non-standard $CP$ transformation
and under the horizontal symmetry
$K$ allows,
in addition to the terms in $V_\phi$ of Eq.~(\ref{Vphi}),
two more terms:
\ba
V_9    & = & i \lambda_9
\left[ 
\left( \phi_1^\dagger \phi_2 \right) 
\left( \phi_1^\dagger \phi_3 \right) - 
\left( \phi_2^\dagger \phi_1 \right)
\left( \phi_3^\dagger \phi_1 \right)
\right],
\label{V9} \\
V_{10} & = & i \lambda_{10} 
\left( \phi_2^\dagger \phi_3 - \phi_3^\dagger \phi_2 \right) 
\left( \phi_2^\dagger \phi_2 - \phi_3^\dagger \phi_3 \right).
\label{V10} 
\ea
The coupling constants $\lambda_9$ and $\lambda_{10}$ are real,
and now the same holds for $\lambda_8$ in Eq.~(\ref{Vphi}),
i.e.\ $\epsilon \equiv \arg \lambda_8 = 0$ or $\pi$.
As for the soft breaking of
$K$,
besides $V_\mathrm{soft}$, the $CP$-invariant term
\be
V'_\mathrm{soft} = i \mu'_\mathrm{soft}
\left( \phi_2^\dagger \phi_3 - \phi_3^\dagger \phi_2 \right)
\label{soft'}
\ee
is now allowed too.
Thus the full potential is 
$V = V_\phi + V_9 + V_{10} + V_\mathrm{soft} + V'_\mathrm{soft}$. 
The VEVs of the additional terms in the potential are given by
\ba
\left\langle 0 \left| V_9 \right| 0 \right\rangle &=&
- \lambda_9 u_1^2 u^2 \sin{2\sigma} \sin{\left( \alpha + \beta \right)}\, , \\
\left\langle 0 \left| V_{10} \right| 0 \right\rangle &=& 
\lambda_{10} u^4 \cos{2\sigma} \sin{2 \sigma}
\sin{\left( \alpha - \beta \right)}\, , \\
\left\langle 0 \left| V'_\mathrm{soft} \right| 0 \right\rangle
& = & \mu'_\mathrm{soft} u^2 \sin{2\sigma}
\sin{\left( \alpha - \beta \right)}\, .
\ea

To find the minimum of the potential,
we proceed in analogy to Section~\ref{Z2model}.
We define a function $\mathcal{F}_\phi = F_\phi + 
\left\langle 0 \left| V_9 \right| 0 \right\rangle +
\left\langle 0 \left| V_{10} \right| 0 \right\rangle$
and search for its minimum with respect to $\sigma$,
$\alpha$,
and $\beta$.
We find
\be
\sigma = \frac{\pi}{4}\, , \quad \alpha = \beta = \omega\, ,
\label{minCP}
\ee
which agrees with the minimum of $V_\phi$ in Eq.~(\ref{min}),
except that the value of $\omega$ is now determined by
\be
2 \lambda_8 \sin{2\omega} + \lambda_9 \cos{2\omega} = 0\, .
\label{omega}
\ee

Switching on the two soft
$K$-breaking terms,
the minimum in Eq.~(\ref{minCP}) gets shifted.
Because of the complication with the three additional terms in $V$,
we are unable to give an exact solution for the shifted minimum,
like we did in Section~\ref{Z2model}.
We therefore resort to a discussion of the shift of the minimum
to first order in the small quantities $\mu_\mathrm{soft}$
and $\mu'_\mathrm{soft}$. 
We define $\delta_0$,
$\delta_+$,
and $\delta_-$ as characterizing the deviations of $\sigma$,
$\alpha$,
and $\beta$ from their values in Eq.~(\ref{minCP}):
\be
\sigma = \frac{\pi}{4} - \frac{\delta_0}{2}\, , \quad
\alpha = \omega + \delta_+ + \frac{\delta_-}{2}\, , \quad
\beta = \omega + \delta_+ - \frac{\delta_-}{2}\, .
\label{variables}
\ee
We expand the function $\mathcal{F}_\phi$ to second order,
and the soft-breaking terms to first order,
in these small variables.
Dropping the constant terms,
we obtain the expansion
\be
\frac{1}{2} \sum_{a,b}  \mathcal{F}_{ab} \delta_a \delta_b 
+ \sum_a f_a \delta_a\, ,
\label{expansion}
\ee
where $\mathcal{F} \equiv (\mathcal{F}_{ab})$
is the symmetric and positive matrix
of the second derivatives of $\mathcal{F}_\phi$
at the minimum (\ref{minCP}):
\be\label{F}
\begin{array}{rcl}
\mathcal{F}_{00} &=& - \frac{1}{2}\, \tilde \lambda u^4
+ \lambda_9 u_1^2 u^2 \sin{2 \omega}\, , \\
\mathcal{F}_{++} &=& \left( - 8 \lambda_8 \cos{2\omega}
+ 4 \lambda_9 \sin{2\omega} \right) u_1^2 u^2\, , \\
\mathcal{F}_{--} &=&
- 2 \lambda_7 u^4 - 2 \lambda_8 u_1^2 u^2 \cos{2\omega}\, , \\
\mathcal{F}_{0-} &=& - 2 \lambda_8 u_1^2 u^2 \sin{2\omega}
+ \lambda_{10} u^4\, , \\
\mathcal{F}_{0+} &=& 0\, , \\
\mathcal{F}_{+-} &=& 0\, .
\end{array}
\ee
The second term in Eq.~(\ref{expansion}) represents the expansion
of the soft-breaking terms,
with
\be
f_0 = \mu_\mathrm{soft} u^2\, , \quad 
f_+ = 0\, , \quad 
f_- = \mu'_\mathrm{soft} u^2\, .
\label{f}
\ee
Let us now define the vectors $\mathbf{f} = \left( f_0, f_+, f_- \right)^T$
and $\mbox{\boldmath$\delta$} = \left( \delta_0, \delta_+, \delta_- \right)^T$.
The minimum of $V$ is given,
to first order in the $\delta_a$ and in the soft-breaking parameters,
by
\be
\mbox{\boldmath$\delta$} = - \mathcal{F}^{-1} \mathbf{f}\, .
\label{shift}
\ee
Using Eqs.~(\ref{shift}),
(\ref{F}),
and (\ref{f}) it is straightforward to calculate the shifts $\delta_a$.
From the zeros in $\mathcal{F}$ and $\mathbf{f}$
one immediately concludes that $\delta_+ = 0$,
i.e.\ the quantity $\alpha + \beta$ remains $2\omega$
to first order in the small ratios $\mu_\mathrm{soft}/u_1^2$,
$\mu'_\mathrm{soft}/u_1^2$,
$\mu_\mathrm{soft}/u^2$,
and $\mu'_\mathrm{soft}/u^2$.
The ratio of the muon mass to the tau mass is expressed as 
\be
\frac{m_\mu}{m_\tau} = \left| \frac{v_2 - v_3}{v_2 + v_3} \right|
\simeq \frac{1}{2} \left| \delta_0 + i \delta_- \right|,
\ee
showing that a non-zero muon mass is generated
not only by the different absolute values of $v_2$ and $v_3$,
as in Section~\ref{Z2model},
but also by the different phases of those VEVs.

\section{Summary} \label{summary}

In this paper we have shown that the models of Refs.~\cite{GL01,GL03,GLcp},
which achieve maximal atmospheric neutrino mixing
through non-abelian symmetries,
allow for a technically natural explanation
of the smallness of the muon mass as compared to the tau mass.

The models are characterized by diagonal Yukawa couplings, 
as seen in the Lagrangian of Eq.~(\ref{L}),
valid for the models of Refs.~\cite{GL01,GL03},
and in the Lagrangian of Eq.~(\ref{L'}),
in the case of Ref.~\cite{GLcp}. 
At face value those Yukawa Lagrangians suggest that $m_\mu$
and $m_\tau$ should be of the same order of magnitude.
In this paper,
the key for $m_\mu \ll m_\tau$
was the extra horizontal symmetry
$K$ of Eq.~(\ref{T}).
As a consequence of that symmetry
the ratio $m_\mu/m_\tau$ is given by Eq.~(\ref{ratio}),
which is a function only of the VEVs $v_2$ and $v_3$.
We have shown that the invariance of the Higgs potentials under
$K$ and under the additional symmetries of the various models
is such that they all admit minima with $v_2 = v_3$,
leading to $m_\mu = 0$.
We then break
$K$ softly by the term in Eq.~(\ref{soft}),
which is unique in the cases of the $\mathbbm{Z}_2$ model of Ref.~\cite{GL01}
and of the $D_4$ model of Ref.~\cite{GL03};
in the case of the $CP$ model of Ref.~\cite{GLcp} there is the additional
$K$ soft-breaking term of Eq.~(\ref{soft'}).
In this way,
we link the smallness of $m_\mu/m_\tau$
to the smallness of the soft-breaking terms. 

As for the electron mass,
it can be read off from the Yukawa Lagrangians (\ref{L}) and (\ref{L'})
that the smallness of $m_e$
is linked to the smallness of the neutrino masses
through the small VEV of $\phi_1^0$. 

We feel that seesaw models where neutrino mixing stems solely
from the Majorana mass matrix of the heavy neutrino singlets
are quite appealing,
since they allow for symmetries
that enforce maximal atmospheric neutrino mixing.
It is,
therefore,
possible that our mechanism for $m_\mu \ll m_\tau$,
which is connected with Yukawa Lagrangians
of the type in Eqs.~(\ref{L}) and (\ref{L'})
and does not need any new fields,
has a wider applicability 
and is not confined to the models of Refs.~\cite{GL01,GL03,GLcp} 
discussed in this paper.

\vspace*{10mm}

\noindent \textbf{Acknowledgement} The work of L.L.\ was supported
by the Portuguese \textit{Funda\c c\~ao para a Ci\^encia e a Tecnologia}
under the contract CFIF--Plurianual.


\newpage

\begin{thebibliography}{9}

\bibitem{reviews}
W. Grimus,
\textit{Neutrino physics---Theory},
lectures given at the
\textit{41. Internationale Universit\"atswochen f\"ur Theoretische Physik,
Flavour Physics},
Schladming, Styria, Austria, 22--28 February 2003
[hep-ph/0307149]; \\
J.W.F. Valle,
\textit{Neutrino masses twenty-five years later}, 
invited talk presented at MRST'03, Syracuse, New York, May 2003
[hep-ph/0307192]; \\
V. Barger, D. Marfatia, and K. Whisnant, 
\textit{Progress in the physics of massive neutrinos},
hep-ph/0308123.

\bibitem{seesaw}
M. Gell-Mann, P. Ramond, and R. Slansky, 
\textit{Complex spinors and unified theories},
in \textit{Supergravity, Proceedings of the Workshop,
Stony Brook, New York, 1979},
eds. P. van Nieuwenhuizen and D.Z. Freedman
(North Holland, Amsterdam, 1979); \\
T. Yanagida,
\textit{Horizontal gauge symmetry and masses of neutrinos},
in \textit{Proceedings of the Workshop on Unified Theories
and Baryon Number in the Universe},
Tsukuba, Japan, 1979,
eds. O. Sawada and A. Sugamoto
(KEK report no. 79--18, Tsukuba, 1979); \\ 
R.N. Mohapatra and G. Senjanovi\'c,
\textit{Neutrino mass and spontaneous parity violation},
Phys. Rev. Lett. 44 (1980) 912.

\bibitem{ma}
K.S. Babu, E. Ma, and J.W.F. Valle,
\textit{Underlying $A_4$ symmetry for the neutrino mass matrix
and the quark mixing matrix},
Phys. Lett. B 552 (2003) 207
[hep-ph/0206292];
\\
see also E. Ma,
\textit{Plato's fire and the neutrino mass matrix},
Mod. Phys. Lett. A 17 (2002) 2361
[hep-ph/0211393].

\bibitem{GL01}
W. Grimus and L. Lavoura,
\textit{Softly broken lepton numbers and maximal neutrino mixing},
J. High Energy Phys. 07 (2001) 045
[hep-ph/0105212];
\\
W. Grimus and L. Lavoura,
\textit{Softly broken lepton numbers: an approach to maximal neutrino mixing},
Acta Phys. Polon. B 32 (2001) 3719
[hep-ph/0110041].

\bibitem{GL03}
W. Grimus and L. Lavoura,
\textit{A discrete symmetry group for maximal atmospheric neutrino mixing},
Phys. Lett. B 572 (2003) 189 
[hep-ph/0305046 v2].

\bibitem{GLcp}
W. Grimus and L. Lavoura,
\textit{A non-standard $CP$ transformation leading to maximal
atmospheric neutrino mixing},
to be published in Phys. Lett. B
[hep-ph/0305309 v3].

\bibitem{scott}
P.F. Harrison and W.G. Scott,
\textit{$\mu$--$\tau$ reflection symmetry in lepton mixing
and neutrino oscillations}, 
Phys. Lett. B 547 (2002) 219
[hep-ph/0210197].

\bibitem{ecker}
G. Ecker, W. Grimus, and W. Konetschny, 
\textit{Quark mass matrices in left--right symmetric gauge theories},
Nucl. Phys. B 191 (1981) 465;
\\
G. Ecker, W. Grimus, and H. Neufeld,
\textit{Spontaneous $CP$ violation in left--right symmetric gauge theories},
Nucl. Phys. B 247 (1984) 70;
\\
W. Grimus and M.N. Rebelo,
\textit{Automorphisms in gauge theories and the definition of $CP$ and $P$},
Phys. Rept. 281 (1997) 239
[hep-ph/9506272].

\bibitem{seidl}
T. Ohlsson and G. Seidl,
\textit{A flavor symmetry model for bilarge leptonic mixing
and the lepton masses},
Nucl. Phys. B 643 (2002) 247 
[hep-ph/0206087];
\\
G. Seidl,
\textit{Deconstruction and bilarge neutrino mixing},
hep-ph/0301044.

\end{thebibliography}
\end{document}